\documentclass[prl,twocolumn,aps,superscriptaddress,amsmath,amssymb]{revtex4-1}

\usepackage{graphicx}
\usepackage{dcolumn}
\usepackage{bm}

\usepackage{color}
\usepackage[normalem]{ulem}
\usepackage{hyperref}

\begin{document}

\author{Fabio Leoni}
\affiliation{Dipartimento di Fisica, Universit\`a degli Studi di Roma La Sapienza, Piazzale Aldo Moro 5, Rome, 00185, Italy}
\email{fabio.leoni@uniroma1.it}
\author{Fausto Martelli}
\affiliation{IBM Research Europe, Hartree Centre, Daresbury WA4 4AD, UK}
\author{John Russo}
\affiliation{Dipartimento di Fisica, Universit\`a degli Studi di Roma La Sapienza, Piazzale Aldo Moro 5, Rome, 00185, Italy}

\title{Correlating ultrastability with fragility and surface mobility in vapor deposited tetrahedral glasses}

\begin{abstract}
Several experiments on molecular and metallic glasses have shown that the ability of vapor deposition to produce ultrastable glasses is correlated with their structural and thermodynamic properties. Here we investigate the vapor deposition of a class of tetrahedral materials (including silicon and water) via molecular dynamics simulations of the generalized Stillinger-Weber potential. By changing a single parameter that controls the local tetrahedrality, we show that the emergence of ultrastable behaviour is correlated with an increase in the fragility of the model. At the same time, while the mobility of the surface compared to the bulk shows only slight changes at low temperature, with increasing the tetrahedrality, it displays a significant enhancement towards the glass transition temperature.
Our results point towards a strong connection between bulk dynamics, surface dynamics and glass-ultrastability ability in this class of materials.
\end{abstract}

\maketitle

\section{Introduction}

Experiments of vapor deposition on a cold substrate kept at an optimal temperature below the glass transition performed using several materials including organic molecular \cite{swallen2007,ediger2017} and metallic glasses \cite{luo2018,ediger2017}, have produced glasses with exceptional stability, well beyond those conventionally obtained by annealing the liquid melt.
The enhanced stability of ultrastable glasses is related to their peculiar chemical, mechanical and thermodynamic properties \cite{ediger2017,rodriguez2022,dalal2013,fullerton2017,sepulveda2014,dalal2012,whitaker2015,rodriguez2015}, but a full understanding of what makes a system ultra-stable is hindered by the fact that many of these properties
depend on the material under consideration. For example, organic molecular glasses such as trehalose or indomethacin \cite{Lyubimov2013,ediger2017,rodriguez2022} form anisotropic structures, while deposited glasses composed of spherically symmetric molecules, such as metallic glasses, are isotropic as shown for example by the radial distribution function in different directions \cite{singh2013,leoni2023a}.
Several criteria for ultra-stability have been proposed. Surface and sub-surface~\cite{swallen2007,Lyubimov2013,lyubimov2015,dalal2015,walters2017,berthier2017,bell2003,ellison2003,reid2016,stevenson2008,samanta2019,ferron2022,leoni2023a} mobility is often considered in experiments as well as in simulations as one of the main factor favouring a stable arrangement of the molecules in the glass.
Bulk dynamics also plays an important role in ultrastability, with several works finding a positive correlation between fragility and stability of deposited glasses \cite{rodriguez2022,chatterjee2024}, including experiments on molecular \cite{nakayama2013} and metallic \cite{yu2013} glasses. At the same time stable deposited glasses of strong glass-formers are known in organic glasses \cite{chua2015}. 

To gain microscopic insight into ultrastable glass formation, several simulation works have investigated numerically the process of vapor deposition \cite{singh2011,Lyubimov2013,lupi2014,lyubimov2015,dalal2015,berthier2017,reid2016,zhang2017,seoane2018,moore2019,bagchi2020,zhang2022,leoni2023a,leoni2023b}. Despite the large gap in accessible timescales, simulations have successfully reproduced essential features of several ultrastable deposited glasses (DG), such as the optimal temperature typically in the range $0.8<T/T_g<0.9$, for both molecular glasses \cite{ediger2017} and for those made of symmetrical components \cite{Lyubimov2013,reid2016,berthier2017,leoni2023a}. From the atomic positions, simulations also allow the study of higher-order correlations~\cite{leoni2023a,leoni2023b}, and have convincingly showed that the structure of ultrastable glasses corresponds to that of  quenched glasses with very small quench rates.

However, the class of materials explored by simulations is still small, and a full understanding of how ultrastable behaviour compares for different materials is limited. To address this point, here we explore the ultrastable behaviour of a class of tetrahedral materials described by the generalized Stillinger-Weber (SW) potential.
Tetrahedral materials, forming a network of strong directional bonds, can show peculiar properties, such as thermodynamic, structural and dynamical anomalies, with respect to those modeled by two-body interaction potentials describing for example simple liquids \cite{russo2022}. 
They include several materials largely employed in the semiconductor and electronics industry such as germanium, silicon or silicon dioxide, as well as carbon and pure water which are of essential importance in several fields of science.
Coarse grained models, explicitly accounting for two- and three-body interactions, have been successfully employed to capture many of the peculiar thermodynamic and structural properties of tetrahedral materials, with particular focus on silicon and water. One of the most widely employed model to study these materials is the generalized Stillinger-Weber (SW) potential \cite{stillinger1985,molinero2006}. For this potential it was shown that many material properties are controlled by the tetrahedral parameter $\lambda$. For example, tuning $\lambda$ has allowed the study of how the tetrahedral-like behavior emerges from the simple-liquid behavior \cite{molinero2006,russo2018,smallenburg2014}.
Also, the fragility of the SW system was shown to depend on $\lambda$, with the system going from Arrhenius to super-Arrhenius with increasing $\lambda$ (with the crossover at $\lambda\simeq 21$) \cite{molinero2006,russo2018}.
\begin{figure*}[!ht]
\begin{center}
\includegraphics[width=16cm]{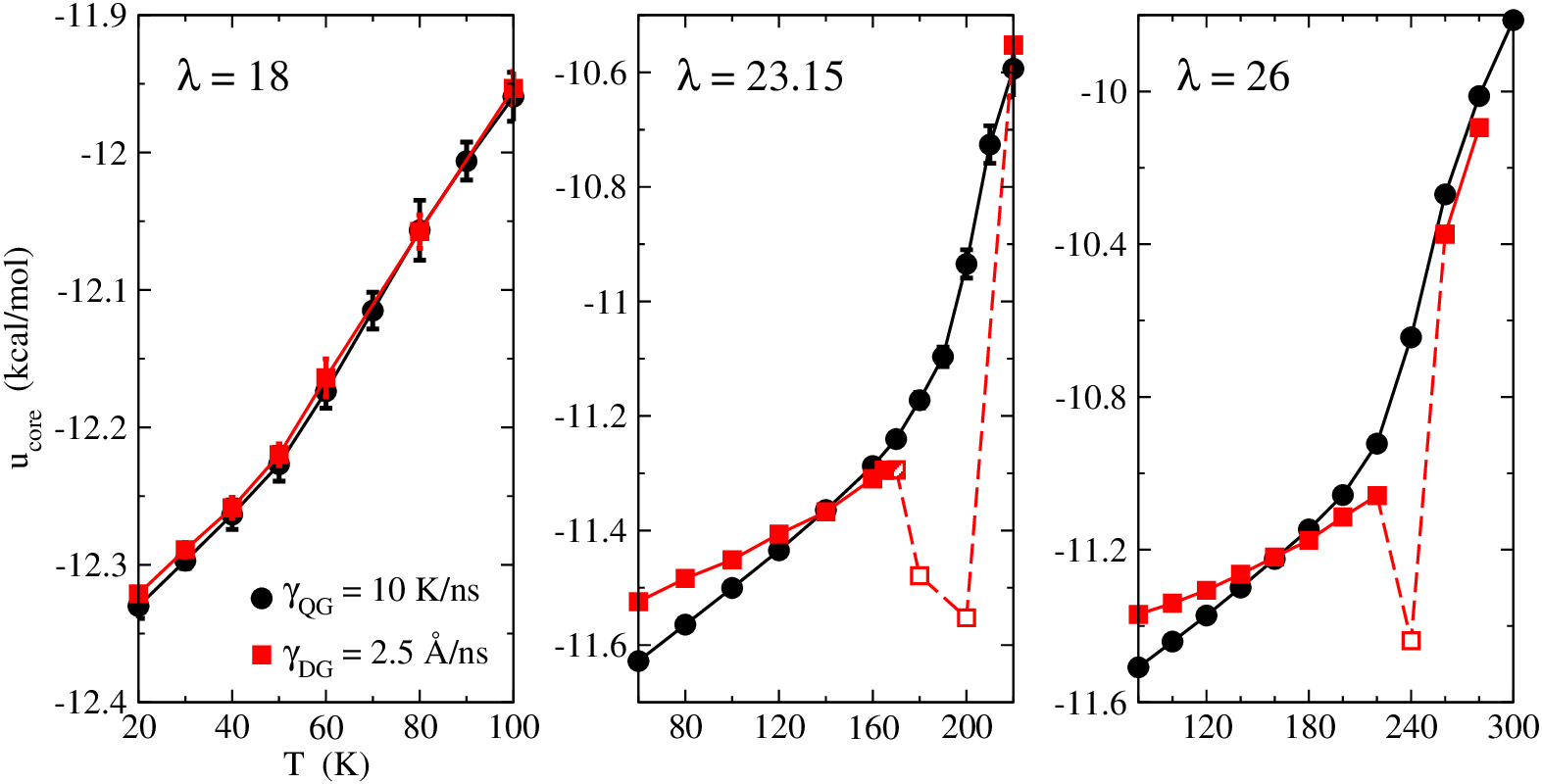}
\caption{\label{fig:u_vs_lambda} Potential energy per particle of the core of the quenched (in black) and deposited (in red) glass layers for $\lambda=18, 23.15$ and 26. Continuous lines connect state points (full symbols) corresponding to glassy phases, while dashed lines connect state points (open symbols) which are partially or fully crystalline. The points are obtained by averaging over 3 different realizations.}
\end{center}
\end{figure*}

In the following, we will investigate for the first time the ability of tetrahedral materials described by the generalized SW potential to form ultrastable glasses by vapor deposition and how their properties depend on their tetrahedral character (controlled by the parameter $\lambda$, see below), with particular attention to the case of water.
Investigating these materials by tuning the tetrahedrality is an excellent opportunity to study how different structural and dynamical properties (such as fragility and surface mobility) affect the ultrastable behaviour (or lack thereof) of the deposited glasses.

\section{Results}

\begin{figure}[!t] 
\begin{center}
\includegraphics[width=8cm]{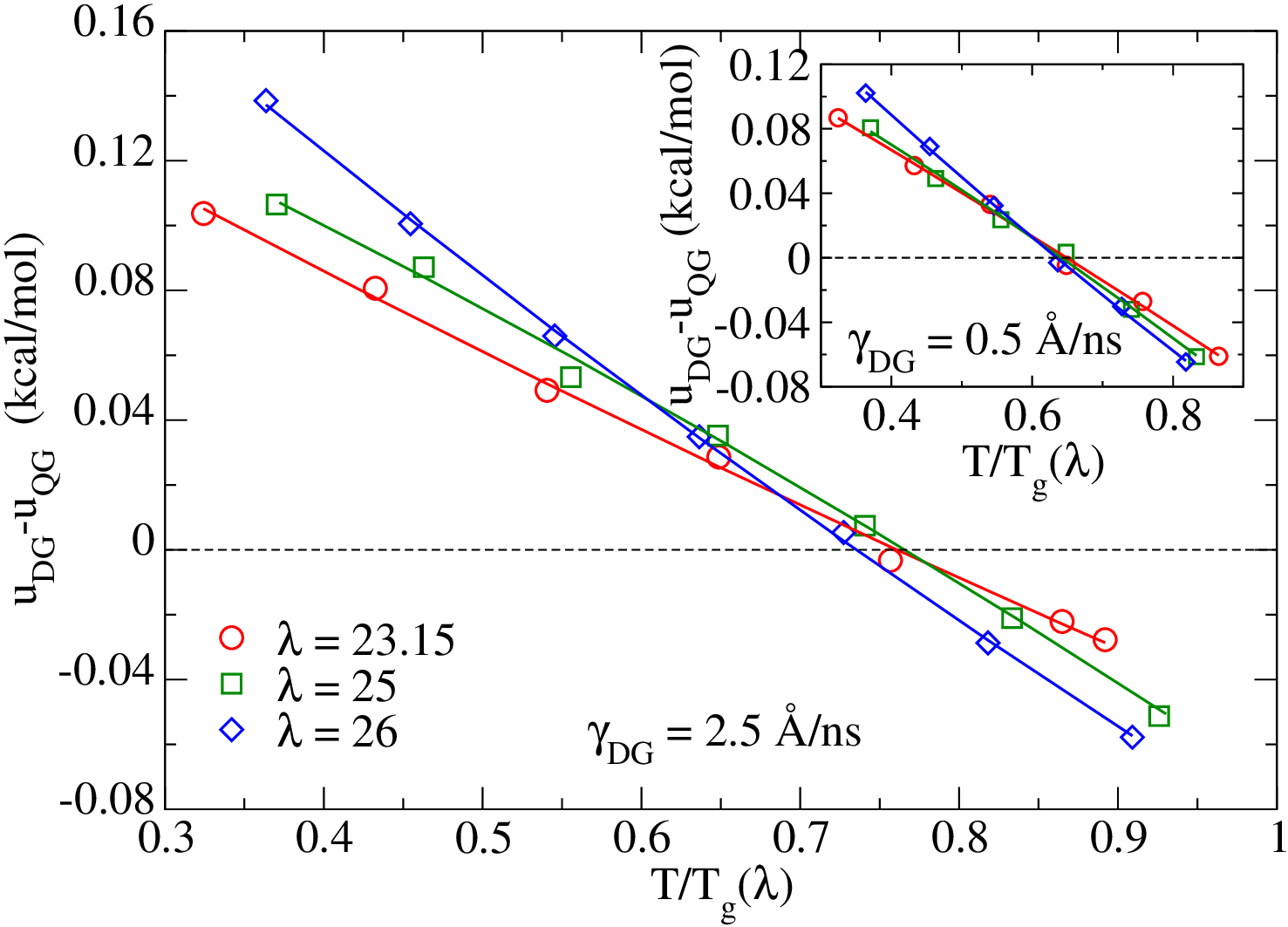}
\caption{\label{fig:uDG_uQG} Core potential energy per particle difference between DG and QG versus scaled temperature $T/T_g(\lambda)$ for $\gamma_{DG}=2.5$~\AA/ns. Symbols are data from Fig.~\ref{fig:u_vs_lambda} and lines are fits. Only values of $\lambda$ for which at least two points in temperature give $u_{DG}<u_{QG}$ are shown. Inset: The same as the main panel, but for $\gamma_{DG}=0.5$~\AA/ns.}
\end{center}
\end{figure}

In the following, we compare vapor-deposited glasses (DG) with quenched glasses (QG) for different values of the tetrahedral parameter $\lambda$.
In Fig.~\ref{fig:u_vs_lambda} we illustrate the thermodynamic stability of the DG and QG systems through the potential energy per particle $u$ as a function of temperature $T$, for $\lambda$ values ranging from 18 to 26. For values of $\lambda$ smaller than 18, crystallization to a body-centered cubic crystal occurs. Hence, the crystalline phase admitted by the system changes from diamond cubic or diamond hexagonal to body-centered cubic \cite{russo2018prx}. The DG obtained for any value of $\gamma_{DG}$ and $T$ always crystallizes during the deposition process. Consequently, we focus our investigation on the formation of the DG for $\lambda\geq 18$.
In Fig.~\ref{fig:u_vs_lambda} we show the potential energy per particle of the core of the layer ($u_{core}$) versus $T$ for the QG (black) and the DG (red) for $\lambda=18, 23.15, 26$. The core of a layer comprises the deposited particles, excluding those within a distance of $5\sigma$ from either the free surface or the interface between the deposited layer and the substrate.
We observe that for $\lambda<21$ the configurations of the QG and the DG correspond to the same value of $u_{core}$ within the error, as shown in the Supplementary Material (SM). On the other hand, by increasing $\lambda$ it is possible to identify a range of temperatures where the DG is more stable than the QG. This range is limited at high temperatures by the nucleation process starting near the glass transition temperature $T_g$. 
At low temperatures, $u_{core}^{QG}$ approaches the value $3/2k_B T$ for any $\lambda$, indicating harmonic behavior. On the other hand, for $\lambda\geq 20$ (see SM) the slope of $u_{core}^{DG}$ suggests the presence of anharmonic components in the energy, as seen also for simple liquids \cite{leoni2023a}. 
Fig.~\ref{fig:uDG_uQG} illustrates the difference in potential energy per particle between the core of the DG and the QG as a function of the scaled temperature $T/T_g(\lambda)$ for $\gamma_{DG}=2.5$~\AA/ns and (Inset) for $\gamma_{DG}=0.5$~\AA/ns. From it we can see that the stability of the DG increases with respect to the QG for both values of $\gamma_{DG}$ as $\lambda$ increases.

\begin{figure}[!th] 
\begin{center}
\includegraphics[width=8cm]{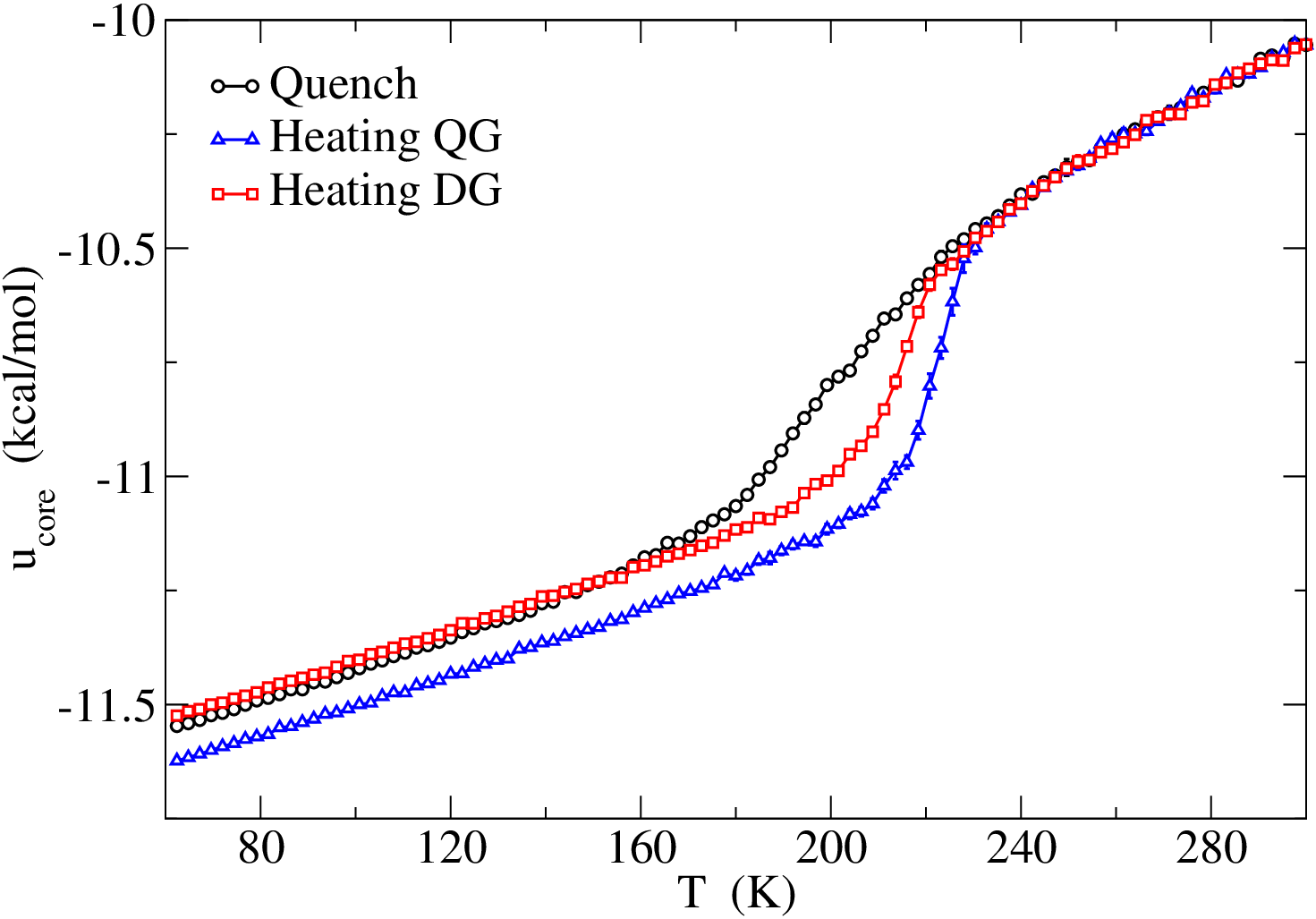}
\includegraphics[width=8cm]{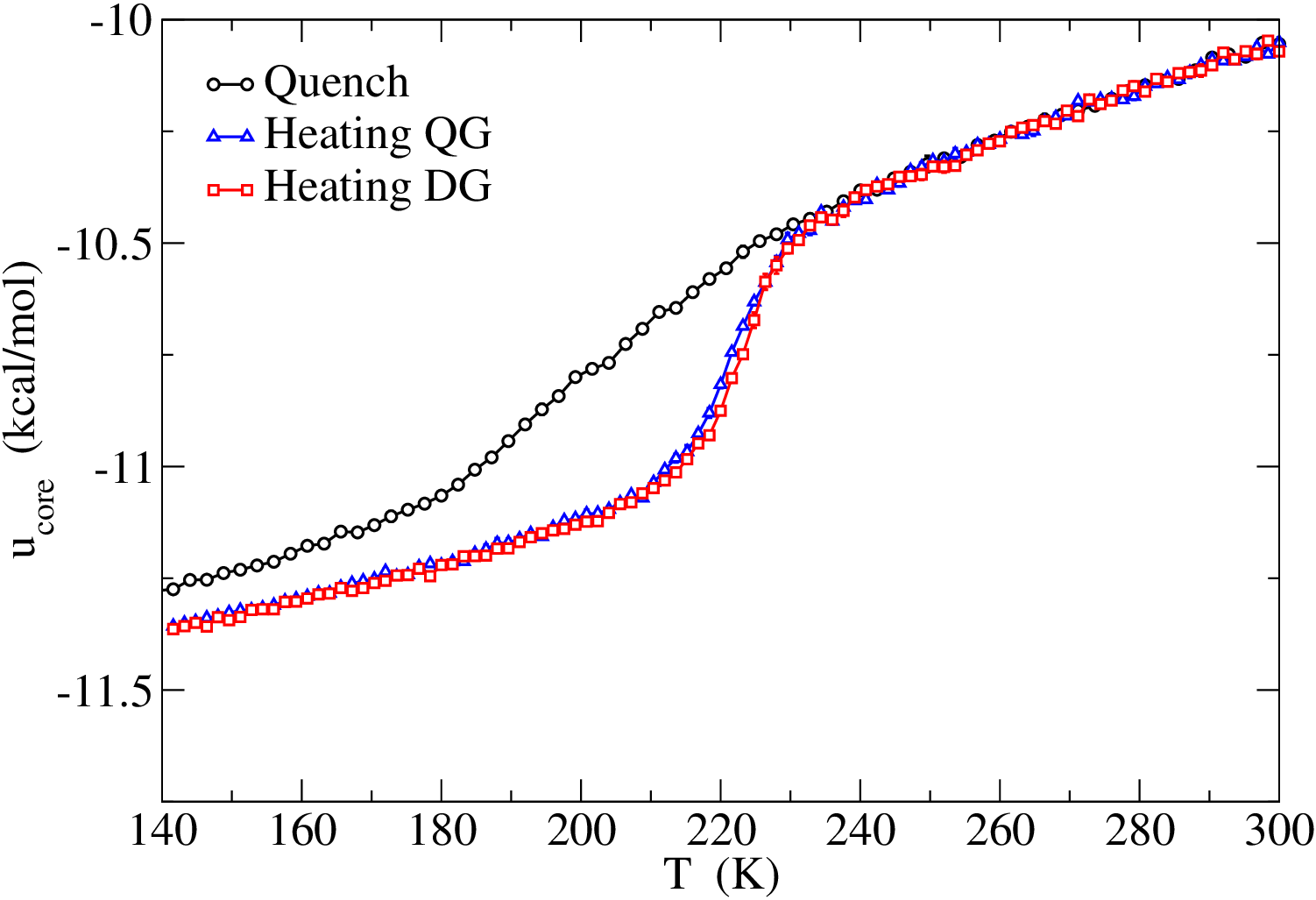}
\includegraphics[width=8cm]{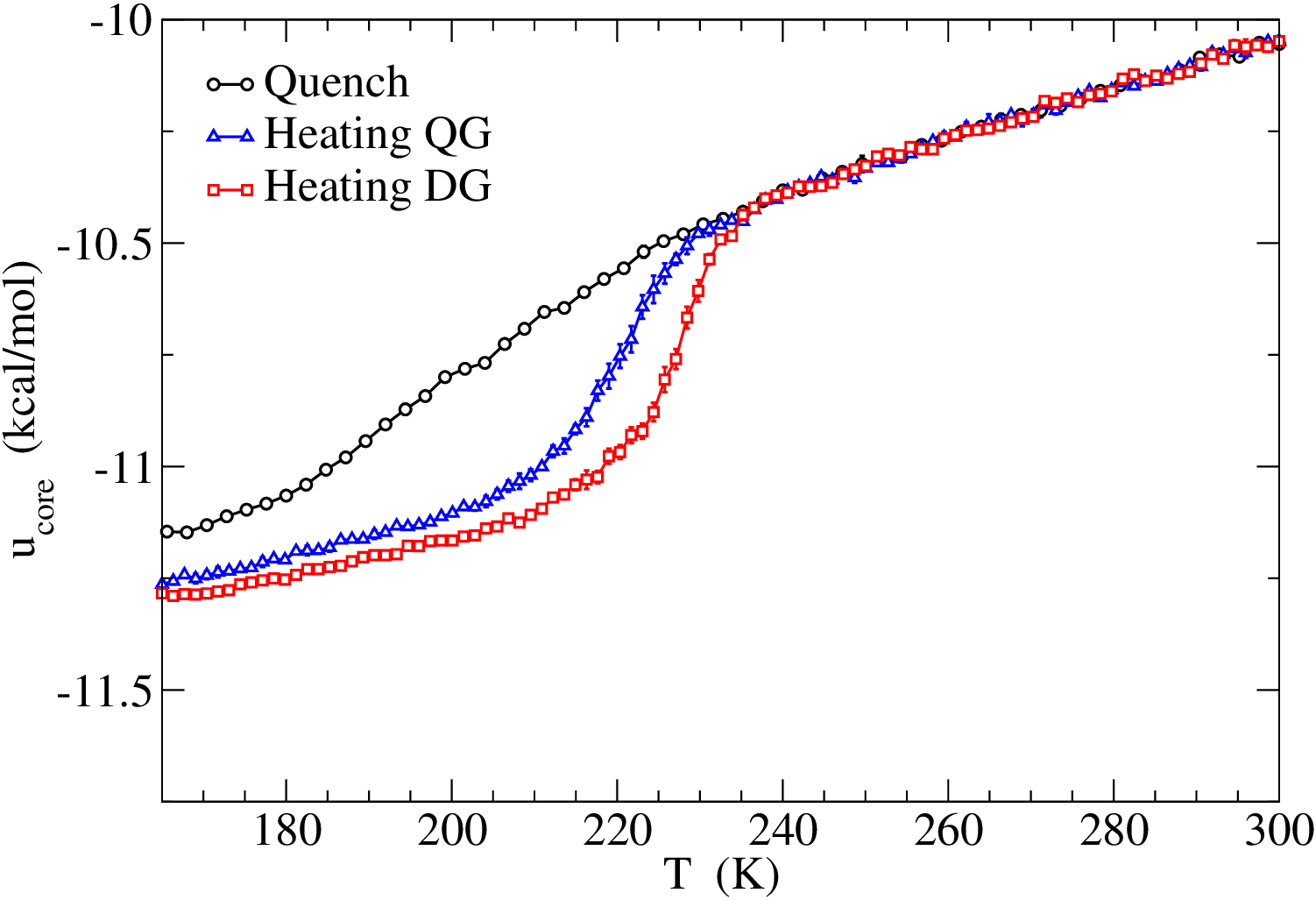}\caption{\label{fig:fast_ramps} Heating and quenching ramps of the DG and QG at rate $100$~K/ns, for temperatures going from $T=60, 140, 165$~K (top to bottom panels) up to $T=300$~K for $\lambda=23.15$.}
\end{center}
\end{figure}

In the SM we show the results of vapor deposition simulations for several values of $\lambda$ and two different deposition rates, confirming the appearance of ultrastable glass formation in the SW system for $\lambda\geq 21$.
We now provide a detailed analysis of the properties of the QG and DG obtained for the case of the monatomic water model mW corresponding to $\lambda=23.15$.
The stability of both the QG and DG is directly related to their respective quench rate, $\gamma_{QG}$, and deposition rate, $\gamma_{DG}$. 
As we decrease $\gamma_{QG}$ and $\gamma_{DG}$, both the QG and DG exhibit enhanced stability until crystallization is reached. In general, decreasing $\gamma_{QG}$, the QG becomes more stable, while for the DG this trend holds true with exceptions \cite{ediger2017}.
\begin{figure}[!t] 
\begin{center}
\includegraphics[width=8cm]{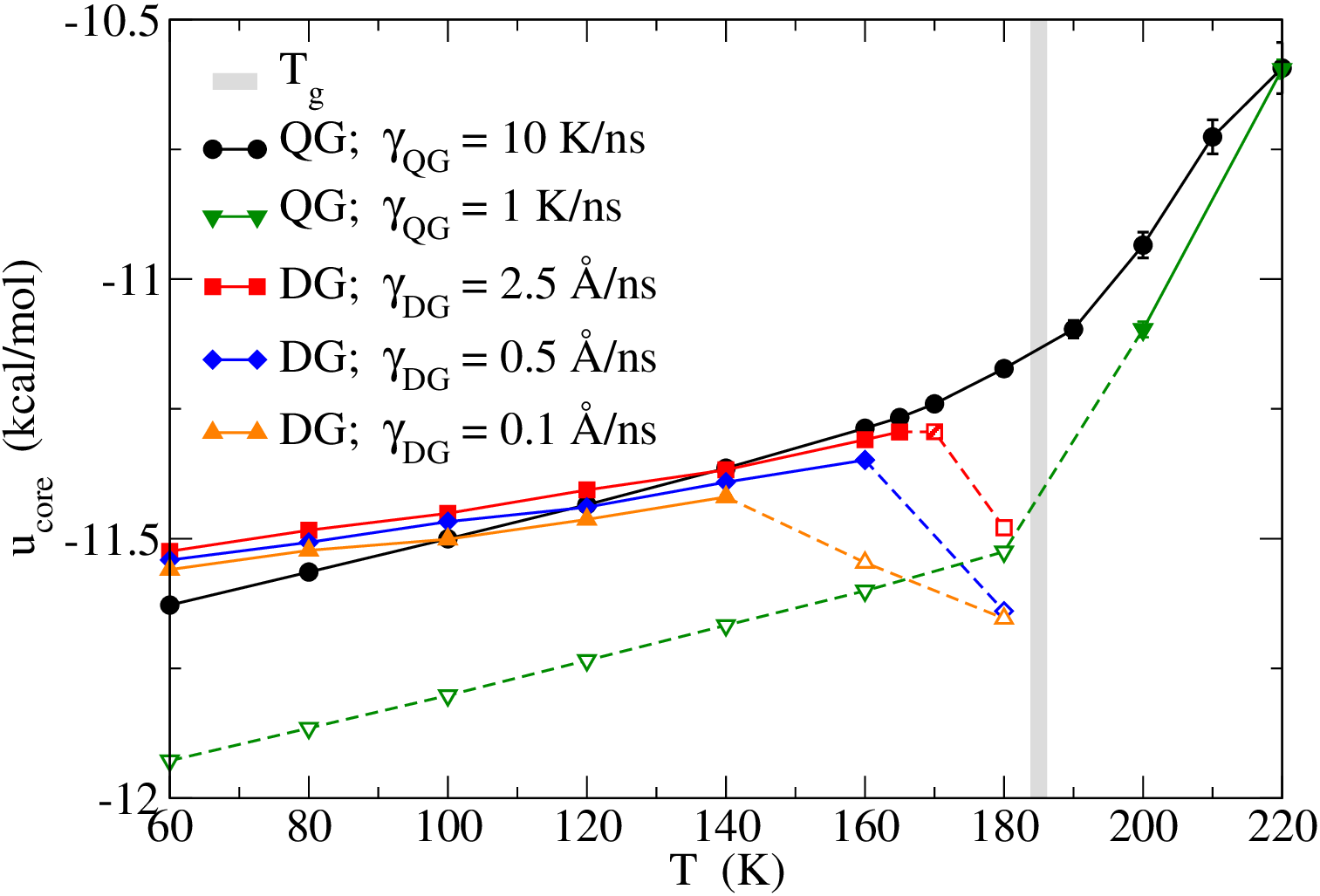}
\caption{\label{fig:u_vs_T_mW} Comparison of the core potential energy per particles $u_{core}$ vs T between the DG and the QG at different deposition rates. All curves are obtained from averaging over 3 different realizations except for DG with $\gamma_{DG}=0.1$~\AA/ns in which case only one realization is considered. The vertical grey band shows the value of $T_g$.}
\end{center}
\end{figure}
In Fig.~\ref{fig:u_vs_T_mW} we show that within the current system, the stability of the DG increases (indicated by a decrease in $u_{core}$) as $\gamma_{DG}$ decreases across all temperature values. However, concurrently, nucleation initiates at lower temperatures. 
Open symbols and dashed lines correspond to the energy of crystallized configurations, while full symbols and continuous lines to that of the supercooled liquid or the glassy phase.
For reference, we show in Fig.~\ref{fig:u_vs_T_mW} $u_{core}$ vs $T$ for the quench rate $\gamma_{QG}=1$~K/ns at which the QG crystallizes.
In the following we compare the QG with the DG at rates corresponding to the integration of an equivalent number of time step per particle (i.e., $5\cdot 10^6$ in the current work), that is $\gamma_{QG}=10$~K/ns for the QG and $\gamma_{DG}=2.5$~\AA/ns for the DG.
The potential energy per particle $u_{core}$ vs $T$ in Fig.~\ref{fig:u_vs_T_mW} suggests that for $\lambda=23.15$, the DG is more stable than the QG within the temperature range $140<T<170$~K.
Crystallization precludes the possibility to compute the structural relaxation time of the DG for temperatures above $T_g\simeq 185$~K. 
Instead, to confirm that the DG is ultrastable with respect to the QG in the range of temperatures $140<T<170$~K, we investigate the kinetic stability of the glass by computing the onset or devitrification temperature ($T_o$) from heating ramps simulations.
Specifically, after a long relaxation of 1 microsecond, we verify that the as-deposited configurations for $T<170$~K are still glassy. 
The onset temperature $T_o$ corresponds to the temperature at which the glass begins to transform into the supercooled liquid. To compute it, we perform rapid heating and quenching ramps at a rate of 100 K/ns such to prevent crystallization (which can occur when heating the glasses at a slower heating rate).
A positive shift in $T_o$ is indicative of an increased kinetic stability \cite{ediger2017,rodriguez2022}.
Figs.~\ref{fig:fast_ramps} shows that at temperatures where the potential energy indicates that the DG is more stable than the QG (i.e., at $T=165$~K, as shown in the lower panel), the $T_o$ of the DG is also larger than that of the QG, and vice versa.
In particular, from the intersection of the glassy and the supercooled lines at $T=165$~K we obtain the results $T_o^{QG}=215$~K and $T_o^{DG}=220$~K indicating the higher stability of the DG with respect to the QG at this temperature.
The ratio $T_o^{DG}/T_g=1.19$ is sensibly larger than 1.05 which is the typical value obtained for the deposition of organic molecules \cite{ediger2017}. However, heating rates used in experiments to probe the onset temperature are considerably lower than the one considered in this simulation work (which is particularly high to avoid crystallization). Therefore, we expect the ratio $T_o^{DG}/T_g$ to decrease for decreasing heating rate.
\begin{figure*}
\begin{center}
\includegraphics[width=8cm]{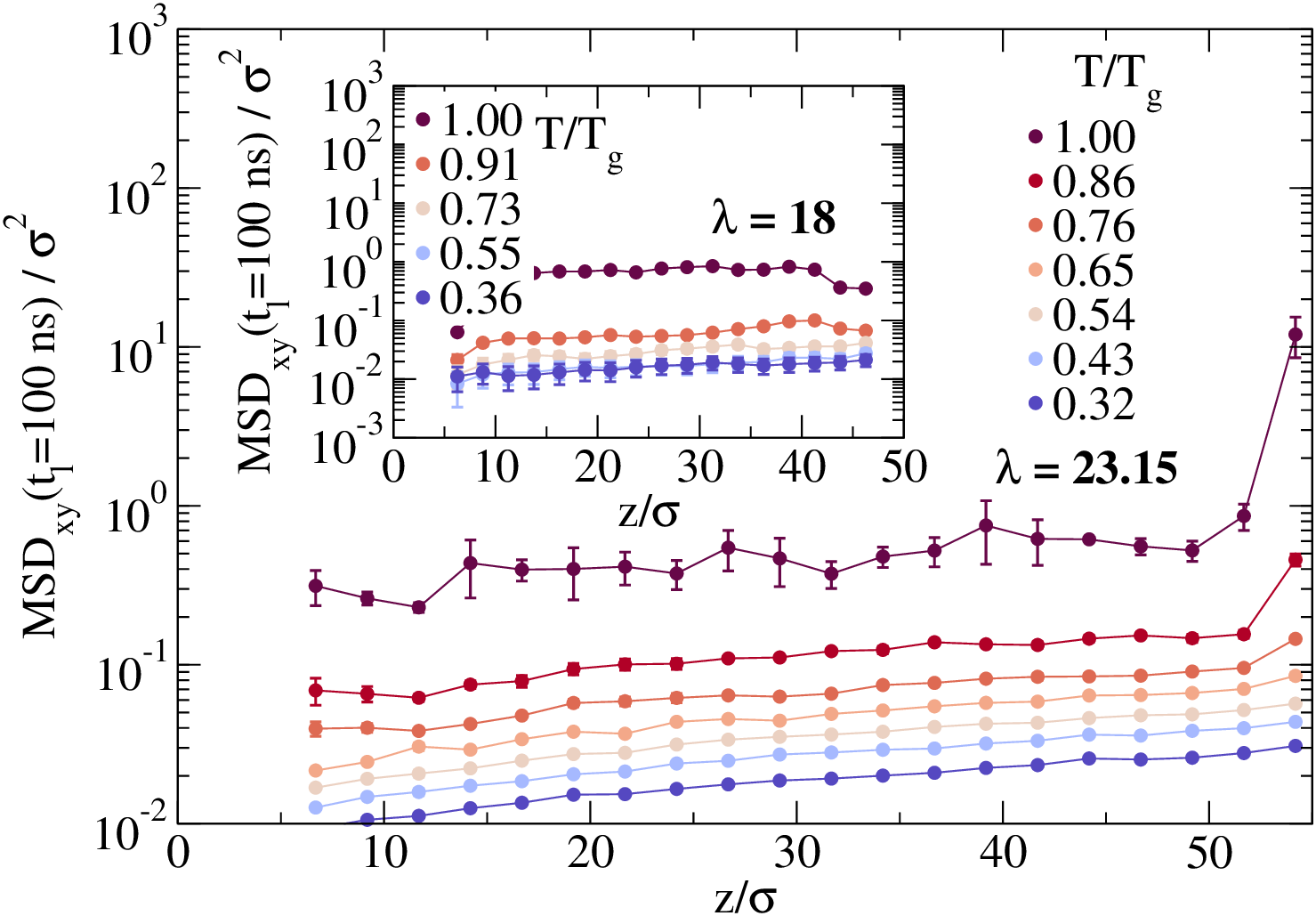}
\includegraphics[width=8cm]{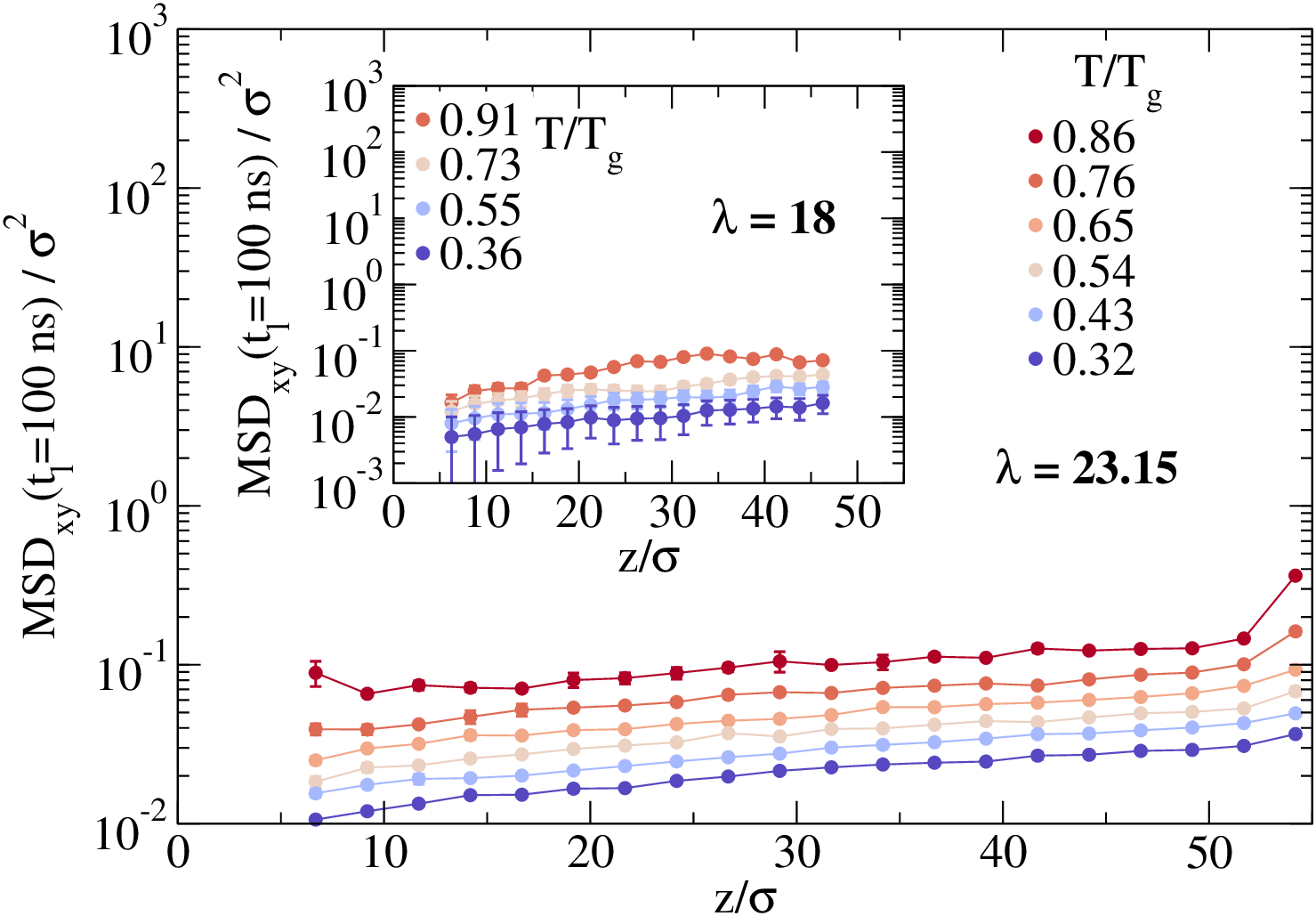}
\caption{\label{fig:msd_10s} Mean square displacement profile (MSD$_{xy}$) along the deposition direction z computed in $2\sigma$-tick xy-slices at the lag time $t_l=100$~ns. Left panel: MSD$_{xy}$ at different temperatures of the QG for $\lambda=23.14$ and (inset) $\lambda=18$. 
Right panel: MSD$_{xy}$ at different temperatures of the DG for $\lambda=23.14$ and (inset) $\lambda=18$.}
\end{center}
\end{figure*}

Another route to quantify the stability of the deposited glass is to compute the ratio between the isothermal transformation time $t_{tr}$, which is the time necessary to transform the glass into a supercooled liquid during isothermal relaxation, and the structural relaxation time $\tau_{\alpha}$ of the system in the bulk, both computed at a temperature $T>T_g$ \cite{tylinski2016}.
We first estimate the isothermal transformation time of the DG for $\lambda=23.15$ obtained at the deposition temperature $T=165$~K by performing relaxation simulations at the fixed temperature of $T=220$~K (i.e., $T/T_g\simeq 1.2$). 
From the mean square displacement (MSD) of the deposited layer we estimate for this system the isothermal transformation time $t_{tr}\simeq 150$~ps.
Therefore, we compute $\tau_{\alpha}$ for the system with $\lambda=23.15$ in the bulk at zero pressure (since glasses prepared in presence of a free surface equilibrate at the pressure of the vapor phase which is close to zero from the thermodynamic point of view) at a temperature of $T=220$~K. 
By fitting the self-intermediate scattering function, computed at the first peak of the static structure factor (corresponding to the wave vector $|q|=2.3$), with the Kolrausch-Williams-Watts law, we obtain the value $\tau_{\alpha}=0.54$~ps. 
Finally, the ratio between the isothermal transformation time and the structural relaxation time for the chosen parameters is found to be $t_{tr}/\tau_{\alpha}\simeq 0.3*10^3$. 
This value compares with those typically obtained for ultrastable deposited glasses through simulations, as for example in the case studied in Ref.~\cite{reid2016}, and can be further enhanced by considering different temperatures, $\lambda$ and deposition rates.

While in simple liquids, the dynamics of the surface of a free-standing \cite{Shi2011} or a supported layer \cite{leoni2023a,leoni2023b} can be orders of magnitude faster than that of the core or bulk of the layer, in the case of tetrahedral materials investigated here, this difference is small at low temperatures, while it grows approaching $T_g$. This is due to the strong directional bond network formed by these materials. 
In Fig.~\ref{fig:msd_10s} we show the mean square displacement profile (MSD$_{xy}(z)$) along the deposition direction $z$ computed in $2\sigma$-tick $xy$-slices at the lag time $t_l=100$~ns obtained from isothermal relaxation simulations. In Fig.~\ref{fig:msd_10s}, the MSD$_{xy}$ is plotted at different temperatures for $\lambda=23.15$ and (inset) $\lambda=18$, for the QG (left panel) and the DG (right panel). If for the QG we can compute the MSD$_{xy}$ at $T/T_g\simeq 1$, for the DG the nucleation of crystallites at that temperature prevents its calculation in the diffusive regime.

\begin{figure}[!t] 
\begin{center}
\includegraphics[width=8cm]{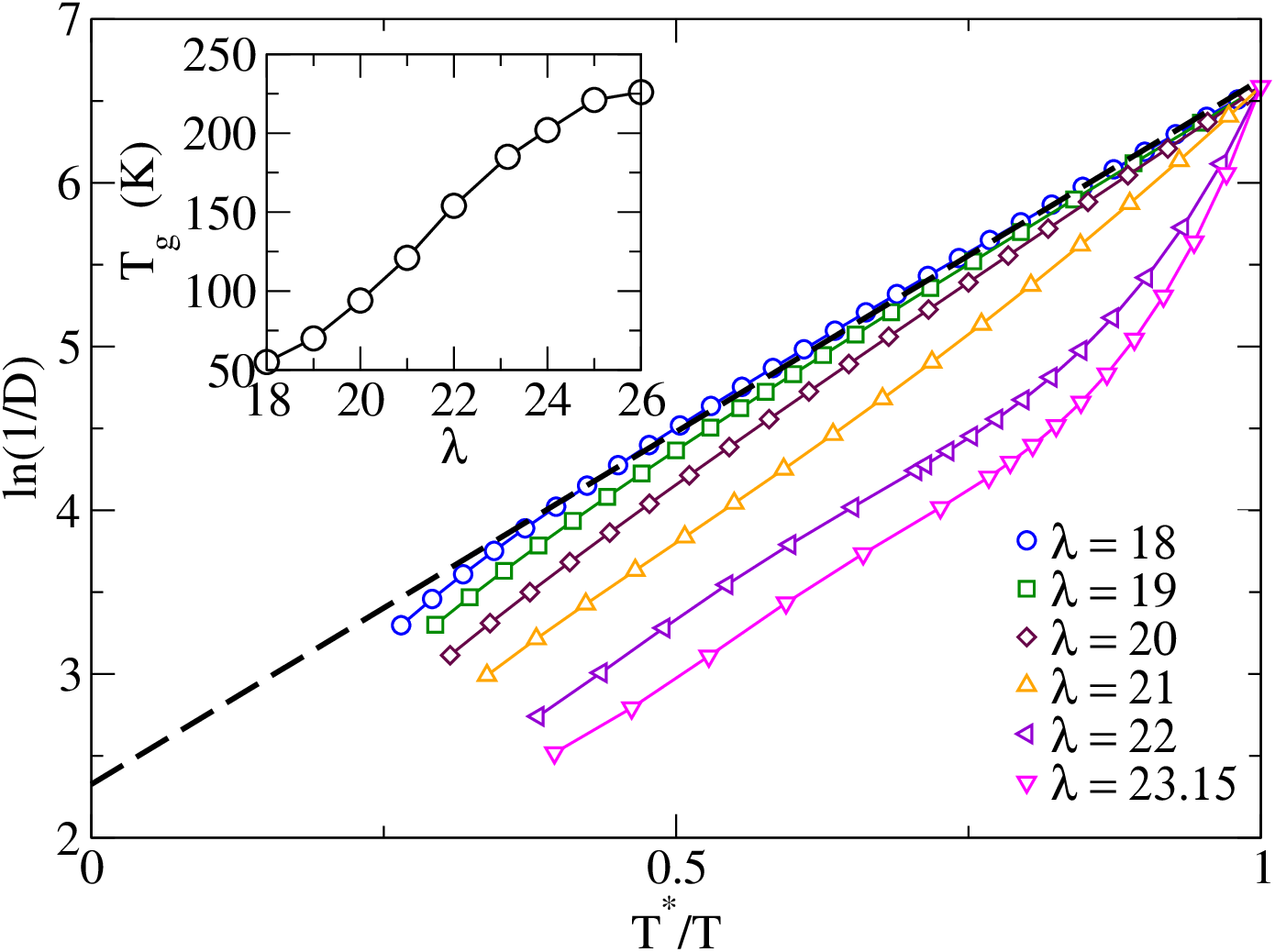}
\caption{\label{fig:Angell} Plot of the logarithm of the inverse of the diffusion coefficient $D$ versus the inverse scaled temperature $T^*/T$ with $T^*\simeq 1.1 T_g$ (see text). Inset: $T_g$ vs $\lambda$ estimated from the intersection between the supercooled and the glassy line from the quench curve of the potential energy.}
\end{center}
\end{figure}
From the MSD profile in Fig.~\ref{fig:msd_10s}, we can see that the mobilities of the QG and DG near the surface are similar at temperatures below $T_g$, irrespective of their stability, as suggested in Ref.~\cite{ZhangY2017}.
As the temperature approaches $T_g$, the mobility at surface is enhanced up to a factor $\sim$~30 for $\lambda=23.15$ (result shown for the QG only due to crystallization of the DG at these conditions).
It has recently been shown \cite{zhang2022} for a deposited glass made of asymmetric molecules that mobility anisotropy at the free surface is associated with the possibility for the glass to lose its stability.
The mobility anisotropy is computed as the ratio between the in-plane (MSD$_{xy}$) and out-of-plane (MSD$_z$) mean squared displacement evaluated at a chosen lag time. 
For the case of $\lambda=23.15$, we obtain a ratio between MSD$_{xy}$ and MSD$_z$ which increases with temperature (for $T/T_g=0.32,...,0.86$) going from $\sim$~3 to $\sim$~4 for the surface, and staying around $\sim$~3 for the core of the DG. Here the surface is the $2.5\sigma$-tick slice close to the free surface, and the core is the $2.5\sigma$-tick slice in the middle of the layer. 
In the case of the QG (which also includes the temperature $T/T_g=1$), for increasing temperature the same ratio goes from $\sim$~3 to $\sim$~22 for the surface, and goes from $\sim$~3 to $\sim$~5 for the core.  
These values suggest that the SW model can potentially form a deposited glass which retains its stability after deposition for temperatures approaching $T_g$, in agreement with the results found in \cite{zhang2022}. We expect this mobility anisotropy to possibly increase for DG obtained below $T_g$ for smaller deposition rates and larger values of $\lambda$.

In the current model system, it is well established that fragility strongly depends on tetrahedrality.
In Fig.~\ref{fig:Angell}, we present an Angell-like plot illustrating the increasing fragility of the system with rising $\lambda$. We plot the logarithm of the inverse of the diffusion constant in the y-axis and the temperatures ratio $T^*/T$ in the x-axis. 
Typically, $T$ is rescaled against $T_g$. However, due to limits in the computational time, we consider the temperature $T^*\simeq 1.1 T_g$ which allows to estimate the diffusion constant for temperatures sufficiently distant from $T_g$ where the dynamics becomes exponentially slower.
Considering the scaled temperature $T^*/T$ 
is sufficient to show the fragility change of the systems with $\lambda$. In particular, the system transitions
from Arrhenius to super-Arrhenius behavior in correspondence of $\lambda\simeq 21$.
In the inset of Fig.~\ref{fig:Angell} we plot $T_g$ vs $\lambda$ as obtained from the intersection between the supercooled and the glassy line of the quench curve 
at the quench rate $\gamma_{QG}=10$~K/ns. It is in good agreement with the estimation of $T_g$ vs $\lambda$ obtained from diffusivity data in Ref.~\cite{russo2018}.
Here we notice that for $\lambda\geq 21$, the SW potential can produce an ultrastable deposited glass which stability increases with $\lambda$, as shown by Fig.~\ref{fig:u_vs_lambda} for two different deposition rates (in the main panel and inset) where the potential energy per particle difference between the DG and QG becomes more negative for increasing $\lambda$ as $T/T_g(\lambda)$ approaches 1. At the same time, we have seen that for $\lambda\geq 21$, the system becomes more fragile for increasing $\lambda$ and the surface mobility increases as well, especially approaching $T_g$ (see Fig.~\ref{fig:msd_10s}). This kind of correlation has been observed in other systems \cite{li2022,chen2016} and it is argued that, since particles (atoms or molecules) at the free surface have in average a lower number of neighboring particles, the effect of the surface, similarly to an increase in temperature, is that of an excitation. On the other hand, fragility controls the rate of increasing of diffusivity with temperature.

\section{Conclusions}
In conclusion, we performed molecular dynamics simulations to investigate the vapor deposition process of tetrahedral materials, including silicon and water, described by the generalized Stillinger-Weber potential. By tuning the tetrahedral parameter $\lambda$ of the potential, we observe that for $\lambda<21$ the conventional quenched glass and the deposited glass exhibit very similar thermodynamic properties. 
On the other hand, increasing the tetrahedral character of the potential, i.e., for $\lambda\geq 21$, we find a range of temperatures where the deposited glass has enhanced thermodynamic properties with respect to the quenched glass. 
These results are obtained by the analysis of the potential energy per particle (probing the thermodynamics stability of the glass) and by the computation of the onset temperature and the ratio between the isothermal transformation time and structural relaxation time (probing the kinetic stability of the glass). The range of temperatures where the deposited glass gains enhanced properties is limited at high temperatures by the nucleation of the crystal phase.
 
We then computed dynamic properties with particular focus on surface mobility which for many materials, especially metallic glasses, is found to be order of magnitudes larger with respect to the bulk, enhancing their stability \cite{leoni2023a}.  
In particular, we compared the dynamics of the surface with respect to the core of the deposited and quenched layers for different values of $\lambda$, finding that surface mobility is significantly enhanced (by a factor of $\sim$ 22 for $\lambda=23.15$) only for the high-tetrahedrality glasses (for $\lambda>21$) and only for temperatures close to the glass transition temperature ($T_g$). Moreover we have shown that the transition from low-tetrahedality glasses to high-tetrahedraility glasses (in correspondence of $\lambda\simeq 21$) is marked by a 
transition from Arrhenius to super-Arrhenius behavior, thus correlating the appearance of ultrasbility with the fragility of the glass.

Our results show that ultrastability can emerge continuously as a function of tetrahedrality. The ability to interpolate continuously between dfferent classes of deposited glasses allows us to compare which factors are responsible for ultrastability. In this respect, our results show that ultrastability is strongly correlated with the appearance of surface mobility around $T_g$, and with super-Arrhenius relaxation in the bulk.


\section{Methods}

The SW potential energy of the system is described by the following expression including two-body ($\phi_2$) and three-body ($\phi_3$) terms:
\begin{equation}
E=\sum_i\sum_{j>i}\phi_2(r_{ij})+\sum_i\sum_{j\ne i}\sum_{k>j}\phi_3(r_{ij},r_{ik},\theta_{ijk})    
\end{equation}
where $r_{ij}$ is the distance between particle $i$ and $j$, and $\theta_{ijk}$ is the angle formed by the triplet of particles $ijk$. The tree-body term $\phi_3$ favors the formation of the tetrahedral angle $\theta_0=109.47^o$ and its strength is controlled by the tetrahedral parameter $\lambda$.
The cutoff is set to $a=1.8\sigma$ with $\sigma$ accounting for the size of a particle.
For the following choice of the parameters (appearing in $\phi_2$ and $\phi_3$), $\lambda=23.15$, $\epsilon=6.189$~kcal/mol and $\sigma=2.3925$~\AA, the SW model reduces to the monatomic water model (mW) \cite{molinero2009} which has been successfully employed to describe ice nucleation \cite{lupi2017,leoni2021} due to its ability to crystallize.
When used, reduced units refer to the mW parameters.

Molecular dynamics (MD) simulations are performed with LAMMPS \cite{LAMMPS}. The simulation box has dimensions $L_x=L_y=15\sigma=35.89$~\AA, while $L_z$ is big enough to accommodate all the particles deposited. Periodic boundary conditions are considered along $x$ and $y$.
The time step is set to $dt=5$~fs, similarly to other deposition simulations of mW particles \cite{lupi2014}.
The deposited glass is obtained by injecting a single particle at a time from a random position at the top of the box over a substrate kept at fixed temperature $T$ (NVT ensemble), as described below.
Particles are injected with a fixed velocity component along the deposition direction, $v_z=0.01$~\AA/fs, and random components along the transversal directions (i. e., along x and y), with $v_x$ and $v_y$ going from -0.001 to 0.001~\AA/fs, corresponding to a source at a temperature of $\sim 1000$~K in the case of the mW.
Injected particles are integrated in the NVE ensemble.
A total of $N=5000$ particles are deposited.
The deposition rate $\gamma_{DG}$ is obtained from the ratio between the size of the deposited layer along the direction $z$ and the total time elapsed: $\gamma_{DG}=\Delta z/\Delta t$. 
Deposition rates considered in the current work are $\gamma_{DG}=0.1, 0.5$ and mainly $2.5$~\AA/ns, which correspond to integrate $5\cdot 10^4$, $10^4$  and  $2\cdot 10^3$ time steps from one injection to the next one, respectively. 

The substrate consisting of 500 particles is obtained by depositing these particles onto a system that contains 500 particles of the same type, arranged randomly.
The random distribution of these particles has a density $\rho=\rho(\lambda,T_m)$, where $T_m(\lambda)$ is the melting temperature at the chosen $\lambda$. To prevent mixing between the deposited particles and those in the random distribution, especially above $T_g$, a spring force of stiffness $k_{rand}=1$ is independently applied to each particle of the random distribution to tether it to its initial position. 
The 500 particles deposited onto the random substrate are introduced at a rate of $\gamma_{sub}=5$~\AA/ns and are given the same velocity vector used for the main deposition simulation.
Subsequently, a substrate of 500 deposited particles is obtained and a spring force with a stiffness of $k_{sub}=1$ is independently applied to each particle, as done for the random distribution.

The conventional quenched glass (QG) that we compare with the vapor-deposited glass (DG) in the following section, is prepared following a similar protocol to that described in Ref.~\cite{leoni2023a}. It consists of taking a DG obtained at the selected value of $\lambda$, melting it at the temperature $T_i(\lambda)>T_m(\lambda)$ and letting the system equilibrate down to $T_f(\lambda)$ at the rate $\gamma_{QG}=\Delta T/\Delta t = (T_i(\lambda)-T_f(\lambda))/\Delta t$. $\Delta t$ is the time to quench the system from $T_i(\lambda)$ to $T_f(\lambda)$ and for the chosen $\gamma_{QG}=10$~K/ns, it corresponds to integrate $5\cdot 10^6$ time steps. The quench rate $\gamma_{QG}=10$~K/ns is fast enough to avoid crystallization in the mW model (corresponding to $\lambda=23.15$).
While in the case of the DG the temperature $T$ is referred to the temperature of the substrate (which is in contact with the thermal bath), for the QG it is associated to the entire system.


\section{acknowledgement}
FL and JR acknowledge support from the European Research Council Grant DLV-759187, partial support by ICSC -- Centro Nazionale di Ricerca in High Performance Computing, Big Data and Quantum Computing, funded by European Union -- NextGenerationEU, and the CINECA award under the ISCRA initiative, for the availability of high-performance computing resources and support.

\bibliography{biblio}

\clearpage
\onecolumngrid

\begin{center}
{\bf\large{Supplementary Material for ``Correlating ultrastability and surface mobility with fragility in vapor deposited tetrahedral glasses''}}
\end{center}

\setcounter{equation}{0}
\setcounter{figure}{0}
\setcounter{table}{0}
\setcounter{section}{0}
\makeatletter
\renewcommand{\theequation}{S\arabic{equation}}
\renewcommand{\thefigure}{S\arabic{figure}}
\renewcommand{\thetable}{S\arabic{table}}
\renewcommand{\thesection}{S\arabic{section}}



\begin{figure}[!hb] 
\begin{center}
\includegraphics[width=17cm]{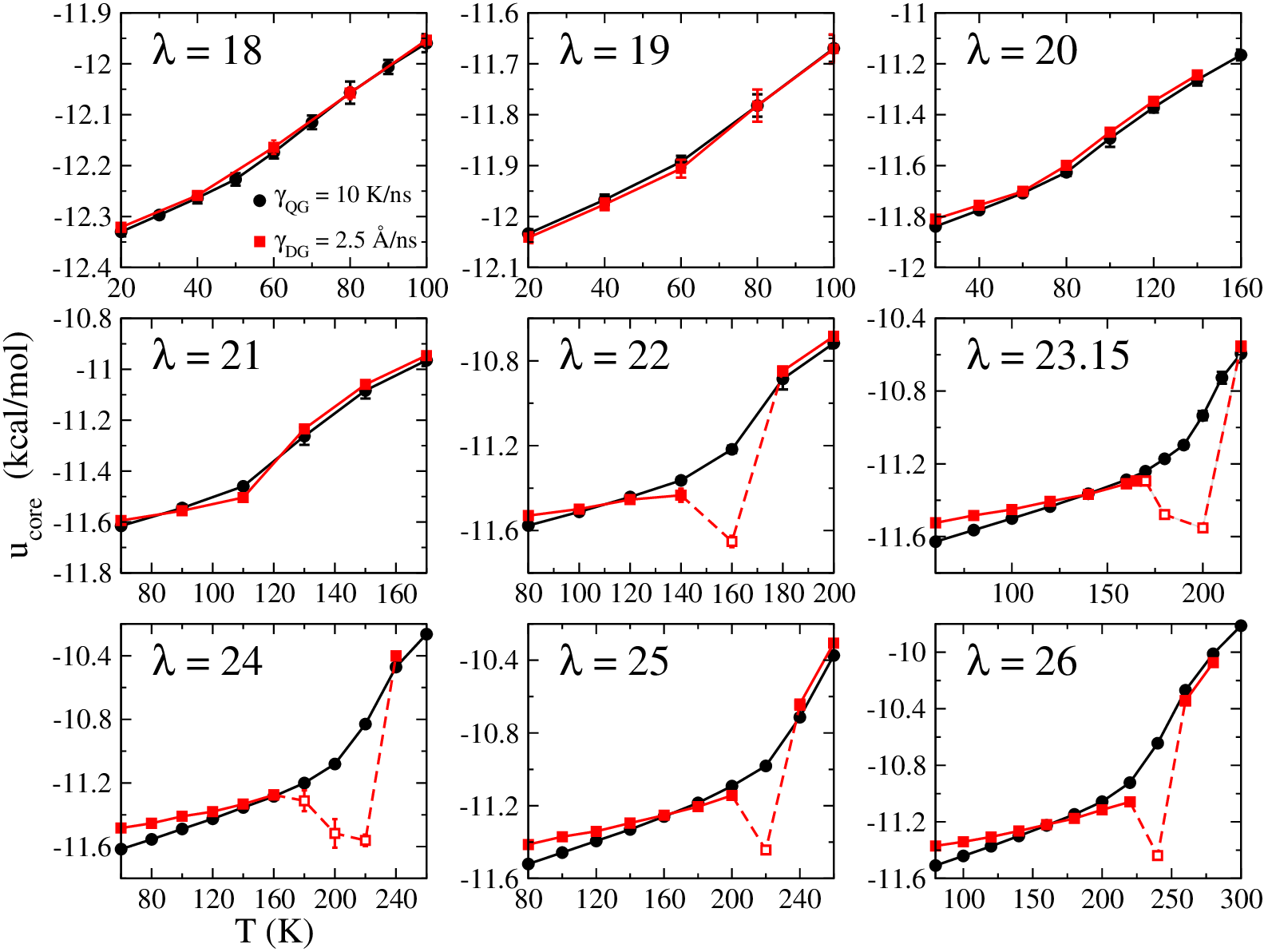}
\caption{\label{fig:} Potential energy per particle of the core of the quenched (in black) and deposited (in red) glass layers for $\lambda=18, 19, 20, 21, 22, 23.15, 24, 25, 26$. The quenched glass is obtained using the quench rate $\gamma_{QG}=10$~K/ns, while the deposited glass using the deposition rate $\gamma_{DG}=2.5$~\AA/ns. Continuous lines connect state points (full symbols) corresponding to glassy phases, while dashed lines connect state points (open symbols) which are partially or fully crystalline. Each point is obtained by averaging over 3 different realizations of quenched or deposited simulations.}
\end{center}
\end{figure}

\begin{figure} 
\begin{center}
\includegraphics[width=17cm]{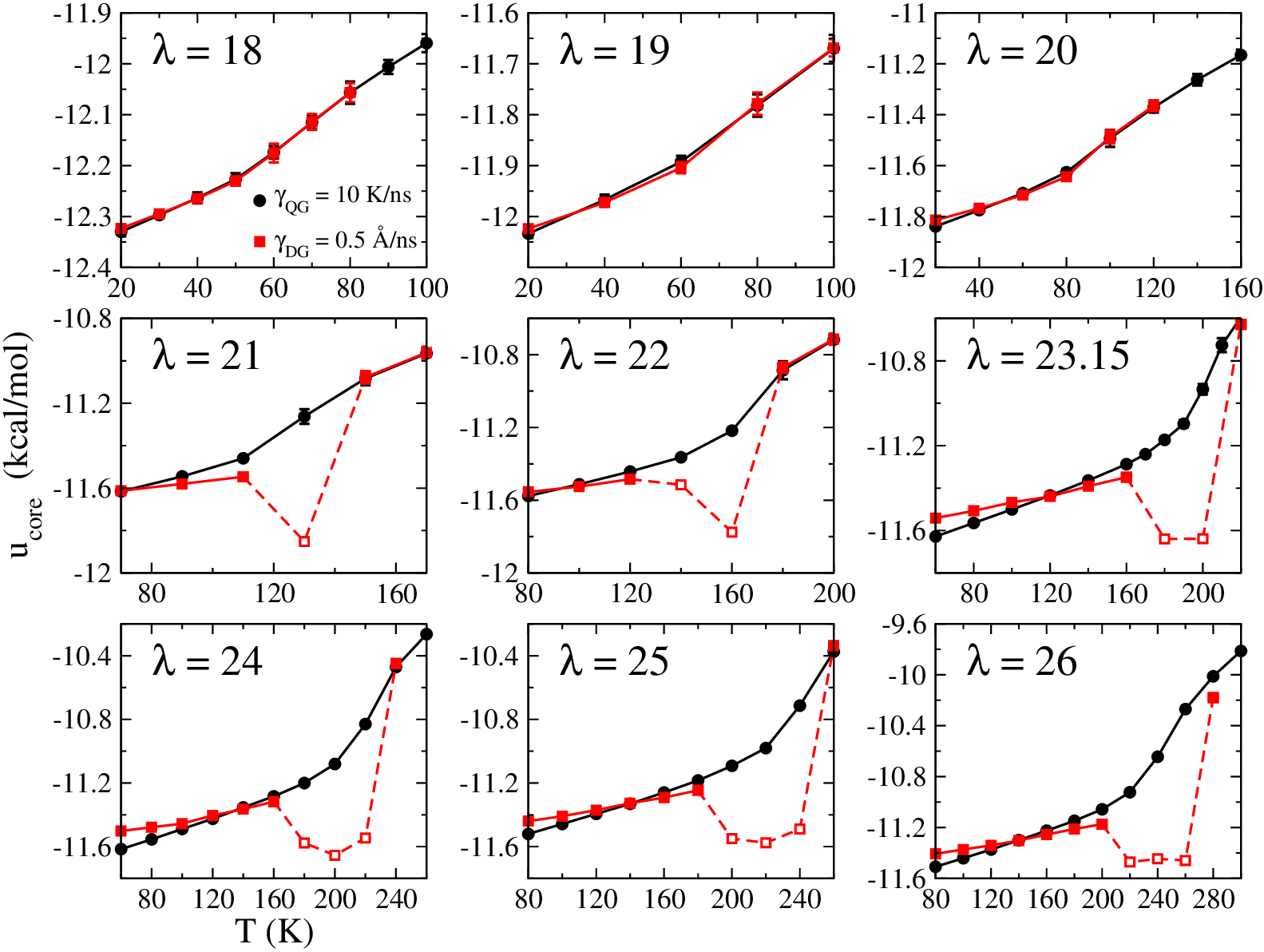}
\caption{\label{fig:} Potential energy per particle of the core of the quenched (in black) and deposited (in red) glass layers for $\lambda=18, 19, 20, 21, 22, 23.15, 24, 25, 26$. The quenched glass is obtained using the quench rate $\gamma_{QG}=10$~K/ns, while the deposited glass using the deposition rate $\gamma_{DG}=0.5$~\AA/ns. Continuous lines connect state points (full symbols) corresponding to glassy phases, while dashed lines connect state points (open symbols) which are partially or fully crystalline. Each point is obtained by averaging over 3 different realizations of quenched or deposited simulations.}
\end{center}
\end{figure}

\end{document}